\documentclass[% 
reprint,
superscriptaddress,
 amsmath,amssymb,
 aps, 
prb,
]{revtex4-2}

\usepackage{graphicx}% Include figure files
\usepackage{dcolumn}% Align table columns on decimal point
\usepackage{bm}% bold math
\usepackage{accents}  % Angstrom \accentset{\circ}{A}
\usepackage{multirow}
\usepackage[version=4,layout=staggered-flat]{mhchem}
\usepackage{diagbox}
\usepackage{makecell}

\begin{document}

\preprint{APS/123-QED}

\title{\textbf{Exact subtraction of electron self-interaction from the Kohn-Sham equations
} 
}% 

\author{J. E. Ingall}
 \email{Contact author: john.ingall@uon.edu.au,\\ john.ingall@alumni.sydney.edu.au}
 \affiliation{Discipline of Physics, The University of Newcastle, Callaghan NSW 2308, Australia}%Lines break automatically or can be forced with \\
\author{Alister J. Page}%
% \email{Contact author: alister.page@newcastle.edu.au}
\affiliation{Discipline of Chemistry, The University of Newcastle, Callaghan NSW 2308, Australia
}%

%\collaboration{CLEO Collaboration}%\noaffiliation

\date{\today}% It is always \today, today,
             %  but any date may be explicitly specified
             
\begin{abstract}
 The self-interaction error (SIE) of Kohn-Sham density functional theory (KS-DFT) leads to a non-physical, non-linear dependence of an orbital's energy on its own fractional occupation [Dabo et al., Phys. Rev. B, 82:115121, 2010]. The {generalized Koopmans condition} ({GKC}) ensures an atomic orbital's eigenenergy is independent of its fractional occupation\textemdash and is free of self-interaction. The constrained search of the Levy spin-density-functional theory yields orbital densities and delivers the ground-state energy to generalize the Hohenberg-Kohn theorems. In this work, the electron density of the Kohn-Sham equations is constrained to be orbital-density dependent, with a total energy functional linear with respect to variation of its orbital densities. That is, the KS multiplicative effective potential $v_{\text{s}}[{n}]$ for an orbital $\varphi_i[{n}]$ is constrained to a functional $v_{\text{eff}}[{n - n_i}]$. The result complies with the generalized Koopmans condition (GKC). This approach improves LDA and GGA mean absolute errors (MAEs) in fundamental band gaps by $\sim$ 1 eV, with respect to KS-DFT. Equations free of electron self-interaction provide the basis for improvements in the accuracy, efficiency, and simplicity of density functional theories.
\end{abstract}

%\keywords{Suggested keywords}%Use showkeys class option if keyword
                              %display desired
\maketitle

%\tableofcontents

\section{\label{sec:intro}Introduction}
% Motivate reader to read this paper
Kohn-Sham density functional theory (KS-DFT) is a first-principles method that reduces a many-body problem to the minimization of a single quantity, \emph{viz.} the energy of a total electron density. KS-DFT is used to calculate a wide range of properties for diverse molecular and material systems with a high level of computational efficiency. The accuracy of any KS-DFT method depends on the approximation of the unknown exchange-correlation interaction between electrons, and the ground- or excited-state property in question. Despite continued progress, challenges remain with the accurate prediction of material properties, in particular spectral and molecular chemical properties including thermochemistry, kinetics and non-covalent interactions \cite{goerigkLookDensityFunctional2017, burschBestPracticeDFTProtocols2022}. Self-interaction errors severely limit the accurate description of large molecular systems with density-functional approximations \cite{broderickDelocalizationErrorPoisons2024, zuluetaFocusDelocalizationError2025, broderickUntanglingSourcesError2025, fanEliminatingDelocalizationError2026}. More generally, delocalization error has been described as the greatest remaining challenge in the development of DFT for the simulation of molecules and materials with high accuracy \cite{bryentonDelocalizationErrorGreatest2023}. Freedom from self-interaction is the primary constraint on a density functional, a condition with which few density functionals comply \cite{jensenIntroductionComputationalChemistry2017}.

% Why the knowledge gap is important, what motivates this study
The predictive accuracy of KS-DFT is increased for exchange-correlation functionals that obey more exact constraints \cite{kaplan_2023}. The {generalized Koopmans condition} ({GKC}) ensures an atomic orbital's eigenenergy is invariant with its own fractional occupation and free of self-interaction \cite{daboKoopmansConditionDensityfunctional2010, nguyenKoopmansCompliantSpectralFunctionals2018}. It has been shown that GKC corrections to solutions of the KS-equations predict highly accurate energies and spectra over a diverse range of archetypal isolated and periodic systems \cite{linscott2023koopmans, schubertPredictingElectronicScreening2024}. For an open system the lowest average energy $E(N)$ is a piecewise-linear (PWL) function of the fractional {\em total} orbital-occupation-number $N$ \cite{perdewSelfinteractionCorrectionDensityfunctional1981}. Deviation from PWL is associated with non-physical delocalization of atomic orbitals \cite{KronikL_2020, cohenInsightsCurrentLimitations2008, mori-sanchezLocalizationDelocalizationErrors2008}. PWL has been generalized to the linearity of a total energy functional with respect to a fractional orbital-occupation-number $f_i \in [0,1]$, for all orbitals $\varphi_i$  \cite{daboKoopmansConditionDensityfunctional2010, schubertPredictingElectronicScreening2024}. This generalized piecewise linearity (GPWL) condition, localized orbitals, and related methods have been found to correct KS-DFT with an accuracy that equals or exceeds that of many-body perturbation theory (MBPT) \cite{daboKoopmansConditionDensityfunctional2010, ferreiraSlaterHalfoccupationTechnique2011, linscott2023koopmans, borghiKoopmanscompliantFunctionalsTheir2014, nguyenKoopmansCompliantSpectralFunctionals2018, degennaroBlochsTheoremOrbitaldensitydependent2022,  borghiVariationalMinimizationOrbitaldensitydependent2015, anisimovTransitionStateMethod2005, kronikExcitationGapsFiniteSized2012, wingBandGapsCrystalline2021, ohadBandGapsHalide2022, kraislerPiecewiseLinearityApproximate2013, maUsingWannierFunctions2016, skoneNonempiricalRangeseparatedHybrid2016, AtallaV_2016,  brawandGeneralizationDielectricDependentHybrid2016, liLocalizedOrbitalScaling2017, Miceli_2018, chenNonempiricalDielectricdependentHybrid2018, yangDensityFunctionalPrediction2020, wengWannierKoopmansMethod2020, maoDFT12Shell2022, ingallAccurateEfficientComputation2024, sagredoElectronicStructureOptical2024, marrazzoSpindependentInteractionsOrbitaldensitydependent2024, schubertPredictingElectronicScreening2024, gouldEnsemblizationDensityFunctional2026, gouldPublishersNoteEnsemblization2026}. These developments followed the recognition that the Hubbard local potential (DFT+U) restored PWL to KS-DFT \cite{cococcioniStudySelectedIron2002, cococcioniLinearResponseApproach2005}. This is also the case for hybrid DFT functionals using a tuned proportion of local and non-local Hartree-Fock (HF) exchange \cite{cohenInsightsCurrentLimitations2008, vydrovTestsFunctionalsSystems2007, hendersonRangeSeparationLocal2008, ingallAccurateEfficientComputation2024}. 

% Why what was done, the problem
If the KS total-energy functional can be constrained to be GPWL, then the aforementioned results suggest there would be no need to remove SIE from self-consistent solutions of the KS equations.
% What was done
Herein we investigate this hypothesis. Potential advantages of this approach are a reduction in computational complexity with a resultant cost similar to KS-DFT. The structure of this manuscript is as follows: Section \ref{sec:background} commences by  briefly re-introducing the standard Levy formalism with generalized KS-equations \cite{levyUniversalVariationalFunctionals1979,hohenbergInhomogeneousElectronGas1964}, and the GPWL and generalized Koopmans conditions (GKC) are then discussed; In Section \ref{s:GKCKS} a single-particle quantum-linearity condition is introduced to constrain the KS effective-potential, which subtracts electron self-interaction in accord with the foregoing conditions.
%the foregoing conditions are applied to orbital-{\em density}-dependent functionals to subtract electron self-interaction from the KS-equations; 
The computational method is described in Section \ref{ss:compdetails}, and computational results are presented in Section \ref{ss:compresults}.

\section{\label{sec:background}Theoretical background}

\subsection{Foundations and formalism}
\label{ss:LLKS}
Let $\Psi$ be the many-particle wave function with orbitals $\psi_{i}({\bf x})$ of a non-relativistic system. For simplicity of the notation ${\bf x} = ({\bf r}, \sigma)$ labels the orbital position ${\bf r}$ and spin $\sigma= \uparrow, \downarrow$, where $\int \mathrm{d} {\bf x}$ implies integration over spatial coordinates and summation over spin coordinates \cite{parrDensityfunctionalTheoryElectronic1995}.

Thomas and Fermi (TF) introduced the concept of an electron gas which is described by the electron density ${n}({\bf r})$ \cite{thomasCalculationAtomicFields1927, fermiMetodoStatisticoDeterminazione1927, liebThomasFermiTheoryAtoms1977}. It was shown by Lenz that the {TF}-equations can be obtained from the minimization of an energy functional that depends only on ${n}({\bf r})$ \cite{lenzUberAnwendbarkeitStatistischen1932, liebThomasFermiTheoryAtoms1977}. The Levy-Lieb (LL) universal functional extends the domain of the Hohenberg-Kohn theory from those electron densities $n({\bf x})$ that determine an external potential $v({\bf x})$, to those densities that normalize to the total electron number $N$ \cite{levyUniversalVariationalFunctionals1979, liebDensityFunctionalsCoulomb1983, hohenbergInhomogeneousElectronGas1964}. In the present work subtraction of orbital self-interaction commences from a generalization of the Levy-Lieb theory to orbital densities. A suitable spin-density-functional theory has been presented by \citeauthor{levyUniversalVariationalFunctionals1979}, \citeauthor{donnellyElementaryPropertiesEnergy1978} \cite{levyUniversalVariationalFunctionals1979, donnellyElementaryPropertiesEnergy1978}, and summarized by \citeauthor{perdewSelfinteractionCorrectionDensityfunctional1981}  \cite{perdewSelfinteractionCorrectionDensityfunctional1981}. The trial density $n$ of the LL functional is generalized to a first-order (reduced) density matrix $\gamma_{i,j}({\bf x}, {\bf x^{'}})$: 
\begin{equation}
	\label{eqn:LevyF}
	Q[\gamma]  \equiv \min_{\Psi \rightarrow {\gamma}} \langle \Psi | \hat{T} + {\hat{V}_{\text{ee}}} | \Psi \rangle ,
\end{equation}
where $\hat{T}$ and $\hat{V}$ are the kinetic and electron-electron repulsion energy operators, respectively. The functional is a constrained search over all $N$-particle, pure-state, antisymmetric wave functions that yield the fixed trial density matrix $\gamma$, to deliver the minimum expectation value. The orbital spin-densities associated with the matrix $\gamma$ are expressed in terms of orthonormal basis wave functions $\varphi_i ({\bf x})$:
\begin{align}
	\label{eqn:rhodef}
	{n}_i({\bf x}) &\equiv \gamma_{ii}({\bf x},{\bf x}) = f_i | \varphi_i ({\bf x})|^2 ,\\
	\label{eqn:rhodef-total}
	{n}({\bf x}) &\equiv \sum_i {n}_i({\bf x}) ,
\end{align}
where $f_i \in [0,1]$ is the occupation number for each single-electron orbital $\varphi_i({\bf x})$. \citeauthor{valoneOnetooneMappingOneparticle1980} has proposed a modification of the Levy constrained search to extend over all statistical mixtures for a trial KS density, rather than the trial density matrix $\gamma$ used here, to give the orbital densities of Eq. (\ref{eqn:rhodef}) \cite{valoneOnetooneMappingOneparticle1980, perdewSelfinteractionCorrectionDensityfunctional1981}. 

The total energy functional is:
\begin{equation}
	\label{eqn:LE}
	E[{n}] = Q[{n}] + \int v({\bf x}) {n}({\bf x}) \mathrm{d} {\bf x} ,
\end{equation} 
for an external  potential $v({\bf x})$. The ground-state energy is given by:
\begin{equation}
	\label{eqn:LE0}
	E_0[{n}] = \min_{{n}} E[{n}] .
\end{equation}
Following \citeauthor{kohnSelfConsistentEquationsIncluding1965} \cite{kohnSelfConsistentEquationsIncluding1965} the total-energy functional defines an (unknown) exchange-correlation energy $E_{\text{xc}}[{n}]$:
\begin{equation}
	\label{eqn:KSEnergy}
	E[{n}] = T_{\text{s}}[{n}] + E_{\text{H}}[{n}] + E_{\text{xc}}[{n}] + \int v({\bf x}){n}({\bf x}) \mathrm{d} ({\bf x}) .
\end{equation}
The ground-state kinetic energy $T_{\text{s}}[{n}_0]$ is  \cite{perdewSelfinteractionCorrectionDensityfunctional1981}: 
\begin{align}
    \label{eqn:LFT}
	T_{\text{s}}[{n_0}]  &\equiv \min_{\Psi \rightarrow {{n_0}}} \langle \Psi | \hat{T} | \Psi \rangle 
    =  \langle \Phi_{0} | \hat{T} | \Phi_0 \rangle  \nonumber \\
    &= \sum_{i}^{N} \int \varphi^*_i({\bf x}) \, \hat{t} \, \varphi_i({\bf x}) \mathrm{d} {\bf x} ,
\end{align}
where $\Phi_0$ is the the ground-state wave function. The second term $E_{\text{H}}[{n}]$ of Eq. (\ref{eqn:KSEnergy}) is the classical Hartree electrostatic self-repulsion of the electron density:
\begin{equation}
	\label{eqn:KSEFH}
	E_{\text{H}}[{n}] = \frac{1}{2} \int \int \frac{{n}({\bf x}){n}({\bf x}{'})}{|{\bf x} - {\bf x}{'}|} \mathrm{d} {\bf x} \mathrm{d} {\bf x}{'} .
\end{equation}

Following \citeauthor{janakProof1978}, \citeauthor{perdewSelfinteractionCorrectionDensityfunctional1981}, minimization of Eq. (\ref{eqn:KSEnergy}) can proceed in two steps: by variation of the orbitals $\varphi_i$ with fixed occupations $f_i$, and vice versa \cite{janakProof1978, perdewSelfinteractionCorrectionDensityfunctional1981}. An infinitesimal variation of the energy functional is constrained to a fixed number of $N$ orthonormal orbitals $\{ \varphi_i \}$ by Lagrange multipliers $\varepsilon_{ij} = \varepsilon^*_{ij}$:
%\begin{widetext}
\begin{align}
	\label{eqn:LFN}
	 &\frac{\delta}{\delta \varphi^*} \left \{ {E}[{n}] 
     - \sum_{ij}^N \varepsilon_{ij} \left [ \int \varphi_i^*[{n}]({\bf x}) \varphi_j[{n}]({\bf x}) \mathrm{d} {\bf x} - \delta_{ij} \right ] \right \} \nonumber \\
  &\qquad = 0 .
\end{align}
%\end{widetext}
For $E[{n}]$ to be a minimum it is a necessary condition that the variation is zero. The self-consistent field (SCF) equations follow  for the variation $\delta \varphi^*$ that is independent of $\varphi$:
\begin{align}
	\label{eqn:eks}
	&\left \{ \hat{t} + v({\bf x}) + v_{\text{H}}[{n}]({\bf x}) + v_{\text{xc}} [{n}] ({\bf x}) \right \} \varphi_i [{n}]({\bf x}) \nonumber \\
   &\qquad = \varepsilon_i [{n}] \varphi_i [{n}]({\bf x}) ,
\end{align}
where the Hermitian $\varepsilon_{ij}$ have been diagonalized with a unitary transformation, and
\begin{align}
	\label{eqn:vHxcdef}
	v_{\text{H}}[{n}]({\bf x}) &\equiv \frac{\delta {{E}}_{\text{H}}[{n}] }{\delta {n}({\bf x})} , \\
	v_{\text{xc}}[{n}]({\bf x}) &\equiv \frac{\delta {{E}}_{\text{xc}}[{n}] }{\delta {n}({\bf x})} ,
\end{align}
are respectively the  Hartree and exchange-correlation potentials. The potentials are local as a consequence of variational minimization with respect to functionals $\varphi^*[{n}]({\bf x})$ of the total electron density ${n}({\bf x})$ \cite{kummelOrbitaldependentDensityFunctionals2008}. The KS-equations are given by Eqs. (\ref{eqn:eks}), the KS effective potential is:
\begin{equation}
	\label{eqn:KSeff}
	v_{\text{s}} [{n}] \equiv v({\bf x}) + v_{\text{H}} [{n}] + v_{\text{xc}} [{n}] .
\end{equation}

%\section{\label{s:theory}Theory and methods}

\subsection{Orbital-density-dependent functionals}
\label{ss:ODD}
The accuracy of local-density and generalized-gradient approximations (LDA, GGA) to the KS exchange-correlation functional has been improved with the development of orbital-dependent functionals, primarily to correct self-interaction error and include derivative discontinuities \cite{kummelOrbitaldependentDensityFunctionals2008}. These functionals include the meta- and hyper-generalized gradient approximations, and hybrid functionals. The KS requirement for a multiplicative effective potential has been relaxed in generalized KS theory (GKS), to yield additional exchange-correlation functionals that describe a partially-interacting system \cite{seidlGeneralizedKohnShamSchemes1996, perdewUnderstandingBandGaps2017}. There is a distinction between KS orbital-dependent functionals (which are functionals of the total electron density), and functionals dependent on orbital densities. 

Examples of the latter orbital-{\em density}-dependent functionals are the Perdew-Zunger (PZ) self-interaction correction (SIC) and Koopmans compliant functionals, which correct KS-DFT for SIE on an orbital-by-orbital basis \cite{perdewSelfinteractionCorrectionDensityfunctional1981, daboKoopmansConditionDensityfunctional2010, linscott2023koopmans}. Hartree-Fock (HF) theory {\em completely} cancels SIE as the direct electrostatic and exchange operators have the same kernel: $1/|{\bf r} - {\bf r}^{'}|$ \cite{perdewSelfinteractionCorrectionDensityfunctional1981}. It has been shown that for a HF model (of two typical atoms) the Hohenberg-Kohn-Sham potential cannot be local \cite{nesbetDoesExactLocal1999}. The PZ-SIC to KS-DFT {\em approximately} cancels SIE for each orbital \cite{perdewSelfinteractionCorrectionDensityfunctional1981}:
\begin{equation}
    \label{eqn:PZ-SIC}
    E^\text{PZ-SIC}[ n_i ] \equiv E_\text{H}[n_i] + E^\text{DFA}_\text{xc}[n_i] ,
\end{equation}
dependent on the density functional approximation (DFA) of the exchange-correlation (xc) energy. The corresponding orbital-density-dependent effective potential is \cite{perdewSelfinteractionCorrectionDensityfunctional1981}:
\begin{align}
\label{eqn:pzsicpot}
v_{\text{eff}}^\text{PZ-SIC} [\{n_i\}] 
  &= v + v_{\text{H}} [{n}] - v_{\text{H}} [n_i] \nonumber \\
  &\qquad + v_{\text{xc}}^\text{DFA} [{n}] - v_{\text{xc}}^\text{DFA} [n_i] .
\end{align}
This correction substantially improves accuracy in the solid state, especially for strongly correlated materials, whereas the results for molecular properties were often less realistic due to static correlation effects \cite{kummelOrbitaldependentDensityFunctionals2008}. Studies of the PZ-SIC results suggest that SIE obtained with semi-local functionals mimics static correlation \cite{cohenChallengesDensityFunctional2012, kummelOrbitaldependentDensityFunctionals2008}. Other generalizations of KS-DFT that address static correlation error include fractional-spin localized orbital scaling correction \cite{suDescribingStrongCorrelation2018}, ensemble DFT \cite{gouldEnsemblizationDensityFunctional2026, gouldPublishersNoteEnsemblization2026, leeRecentAdvancesEnsemble2022}, an ensemble generalization of the PZ-SIC \cite{schwalbeEnsembleGeneralizationPerdew2024} and combined one- and two-electron reduced density matrix functional theory \cite{gibneyUniversalGeneralizationDensity2023, gibneyEnhancingDensityfunctionalTheory2024}. 

The introduction of orbital-density-dependent functionals presented several computational challenges that have been addressed effectively. 
A characteristic of the PZ-SIC is that unitary transformation of occupied orbitals can change the energy even though the total density remains unchanged \cite{kummelOrbitaldependentDensityFunctionals2008}. Similarly, Koopmans functionals add orbital-density-dependent corrections to approximate KS-density functionals, which (as for the PZ-SIC) are not invariant with respect to the unitary transformation of occupied single-particle orbitals \cite{borghiVariationalMinimizationOrbitaldensitydependent2015}. The PZ-SIC also showed numerical instability with multiple local energy minima and the non-physical symmetry breaking of molecular bonds \cite{lehtolaComplexOrbitalsMultiple2016, schwalbeEnsembleGeneralizationPerdew2024, biSelfinteractionCorrectedSCAN2024, richterReducingElectronSelfInteraction2025}. 
An orbital {\em density-potential} constraint has been introduced to achieve efficient convergence of the PZ-SIC equations for molecular and extended systems, leading to improved ionization potentials, molecular dissociation, band gap predictions and the correct treatment of molecular symmetry \cite{biSelfinteractionCorrectedSCAN2024, grimmeR2SCAN3cSwissArmy2021}. An ensemble generalization of the PZ-SIC has also been shown to prevent the spurious breaking of molecular symmetries \cite{schwalbeEnsembleGeneralizationPerdew2024}. 

In contrast to the PZ-SIC, the Koopmans functionals comply with GKC and GPWL conditions. It follows from \citeauthor{janakProof1978}'s theorem \cite{janakProof1978} that the GPWL and GKC conditions are equivalent \cite{daboKoopmansConditionDensityfunctional2010, linscott2023koopmans}. \citeauthor{janakProof1978}'s theorem generalizes KS-DFT from integer to fractional orbital-occupation-numbers:
\begin{equation}
	\label{eqn:Janak}
	\left . \frac{\partial {E}(f_1,f_2,\ldots)}{\partial f_i} \right |_{\bf f} = \varepsilon_i({\bf f}),
\end{equation}
where ${\bf f} = (f_1, f_2, \ldots)$ is a set of the arbitrary and fixed occupation numbers with $f_i \in [ 0,1 ]$, and the $\varepsilon_i$ are ordered energy eigenvalues \cite{dreizlerDensityFunctionalTheory2012}. In general, Eq. (\ref{eqn:Janak}) is non-linear with respect to the variables $\{ f_i \}$. The GPWL condition defines zero curvature of the total energy with respect to the occupation of each orbital \cite{daboKoopmansConditionDensityfunctional2010}:
\begin{equation}
	\label{eqn:GPWL}
	\left . \frac{\partial^2 {E}(f^{'}_1,f^{'}_2,\ldots)}{\partial^2 f^{'}_i} \right |_{\bf f} = 0,
\end{equation}
where the derivatives are performed with respect to the variable $f^{'}_i$ for arbitrary and fixed ${\bf f} = (f_1, f_2, \ldots)$. The generalized Koopmans condition (GKC) originally derived from virtual ionization processes is \cite{schubertPredictingElectronicScreening2024, daboKoopmansConditionDensityfunctional2010,
borghiKoopmanscompliantFunctionalsTheir2014, nguyenKoopmansCompliantSpectralFunctionals2018, degennaroBlochsTheoremOrbitaldensitydependent2022, borghiVariationalMinimizationOrbitaldensitydependent2015}:
\begin{equation}
	\label{eqn:koopmans}
	\left . \frac{\partial \varepsilon_i(f^{'}_1,f^{'}_2,\ldots)}{\partial f^{'}_i} \right |_{\bf f} = 0 .
\end{equation}
 The GKC ensures that the eigenvalues equal the total energy differences, as each energy eigenvalue $\varepsilon_i$ is independent of the orbital occupation $f_i$  \cite{nguyenKoopmansCompliantSpectralFunctionals2018}. It follows from the partial derivative of Eq. (\ref{eqn:Janak}) that Eq. (\ref{eqn:GPWL}) is true {\em if and only if} Eq. (\ref{eqn:koopmans}) is true. That is, that the GPWL and GKC are equivalent conditions. These conditions constrain the total charge density of KS-DFT to a set of orbital-dependent densities. 
 
 % Koopmans functionals have been implemented as the \texttt{koopmans} package \cite{linscott2023koopmans, schubertPredictingElectronicScreening2024}, and as the `kcw.x' program in Quantum ESPRESSO \cite{giannozziAdvancedCapabilitiesMaterials2017, giannozziQUANTUMESPRESSOModular2009}. 
 The Koopmans integral (KI) total-energy functional is defined by:
\begin{align}
    \label{eqn:KIdefn}
    E^\text{KI} &\equiv E^\text{KS} + \sum_i \alpha_i \Pi_i^\text{uKI} ,
\end{align}
where $\alpha_i$ are orbital-dependent screening factors for a linear response, and
\begin{align}
    \label{eqn:uPIdefn}
    \Pi_i^\text{uKI}(f_i) \equiv &- \int_0^{f_i} \langle \varphi_i | H^\text{KS} (s) |  \varphi_i \rangle \;\mathrm{d}s \nonumber \\
    &+ f_i \int_0^1 \langle \varphi_i | H^\text{KS} (s) |  \varphi_i \rangle \;\mathrm{d}s
\end{align}
are KI corrections to the KS-DFT energies for frozen or {\em un-relaxed} orbitals $\varphi_i$ \cite{ColonnaN_2018, borghiKoopmanscompliantFunctionalsTheir2014}. The first term on the right-side of Eq. (\ref{eqn:uPIdefn}) subtracts the non-linear orbital energy. Whereas, the second term adds a linear interpolation of the orbital-energy between integral values of the orbital's occupation. The screening factors $\alpha_i$ have been formally defined so as to revert the un-relaxed Koopmans’ correction to a fully-relaxed state \cite{ColonnaN_2018}:
\begin{equation}
    \label{eqn:alphadefn}
    \alpha_i \equiv \frac{\Pi_i^\text{rKI}}{\Pi_i^\text{uKI}} ,
\end{equation}
where the correction $\Pi_i^\text{rKI}$ for fully-relaxed orbitals is defined in Reference \cite{daboKoopmansConditionDensityfunctional2010}. The screening factors are calculated ab initio from KS ground-state properties in the linear response and from total-energy differences \cite{ColonnaN_2018, nguyenKoopmansCompliantSpectralFunctionals2018}.

\section{\label{s:GKCKS}Quantum-linear density-functional theory}
Consider the single-particle limit of an electron gas. With $n = n_i$, the resulting single KS-equation exhibits non-linear Hartree energy:
\begin{equation}
    \label{eqn:KS-nonlinear}
	E_{\text{H}}[{n_i}] = \frac{1}{2} \int \int \frac{{n_i}({\bf x}){n_i}({\bf x}{'})}{|{\bf x} - {\bf x}{'}|} \mathrm{d} {\bf x} \mathrm{d} {\bf x}{'} \not\equiv 0,
\end{equation}
and an exchange-correlation term:
\begin{equation}
    v_\text{Hxc}[n_i]\varphi_i ,
\end{equation}
which is non-linear with respect to the orbital $\varphi_i$. However, from quantum-mechanics a non-interacting electron satisfies a linear wave-equation. Define the single-particle {\em quantum linearity} (QL) of DFT by the constraint:
\begin{equation}
    \label{eqn:QL}
    E_\text{H}(n_i) = E_\text{x}(n_i) = E_\text{c}(n_i) \equiv 0, 
\end{equation}
for all single-particle orbitals with density $n_i$. This Eq. (\ref{eqn:QL}) is a stronger condition compared with the PZ-SIC \cite{perdewSelfinteractionCorrectionDensityfunctional1981}, as the latter assumes approximate cancellation between the Hartree and xc-energies, Eq. (\ref{eqn:PZ-SIC}). In contrast to the KS-density $n$, we posit the orbital densities $\{ n_i \}$ as approximate independent variables.

We are now prepared to apply the QL condition to remove self-interaction terms from the KS-equations. It is sufficient to commence with integer occupation numbers $f_i \in \{ 0,1 \}$. The Hartree energy functional $E_{\text{H}}[n]$, Eq. (\ref{eqn:KSEFH}), becomes on substituting the total electron density of Eq. (\ref{eqn:rhodef-total}) for non-interacting orbitals $\{ \varphi_i \}$:
\begin{align}
	\label{eqn:KSEF-J-SIE}
	E_{\text{H}}[{n}] &= \frac{1}{2} \int \int \sum_{i=1}^N \frac{{n}_i({\bf x}) {n}({\bf x}{'}) }{|{\bf x} - {\bf x}{'}|}
	 \mathrm{d} {\bf x} \mathrm{d} {\bf x}{'} \\
	&= \frac{1}{2} \sum_{i=1}^N \int \int \frac{{n}_i({\bf x}) {n}({\bf x}{'}) }{|{\bf x} - {\bf x}{'}|}
	\mathrm{d} {\bf x} \mathrm{d} {\bf x}{'} \\
	\label{eqn:KSEF-HSIE}
	&= \frac{1}{2} \sum_{i=1}^N \int \int 
    \left \{ \frac{{n}_i({\bf x})[ {n}({\bf x}{'}) - {n}_i({\bf x}{'} )]}{|{\bf x} - {\bf x}{'}|} \right. \nonumber \\
	&\left. \qquad\qquad\qquad\;\;+ \frac{{n}_i({\bf x}){n}_i({\bf x}{'} )}{|{\bf x} - {\bf x}{'}|} \right \} \mathrm{d} {\bf x} \mathrm{d} {\bf x}{'} \\
    \label{eqn:HSIE-DEFN}
    &\equiv {E}_{\text{H}}[{n}, \{n - n_i\}] + E_{\text{Hsi}}[{n_i}] .
\end{align}
The second term of Eqs. (\ref{eqn:KSEF-HSIE}, \ref{eqn:HSIE-DEFN}) is a Hartree orbital self-interaction energy:
\begin{equation}
\label{eqn:EHsi}
E_{\text{Hsi}}[{n_i}] \equiv \frac{1}{2} \sum_{i=1}^N \int \int 
	\frac{{n}_i({\bf x}){n}_i({\bf x}{'} )}{|{\bf x} - {\bf x}{'}|} 
 \mathrm{d} {\bf x} \mathrm{d} {\bf x}{'} ,
\end{equation}
which for spin-resolved densities $n_i({\bf x}) = n_i({\bf r}\sigma)$ is non-physical. The first term of Eqs. (\ref{eqn:KSEF-HSIE}, \ref{eqn:HSIE-DEFN}) is the Hartree total energy functional:
\begin{align}
    \label{eqn:HTOT}
    {E}_{\text{H}}[{n}, \{n - n_i\}] &= \sum_{i=1}^N {E}_{\text{H}}[{n_i}, n - n_i] ,
\end{align}
where for the single-electron orbital with density $n_i$: 
\begin{equation}
	\label{eqn:HODD-SIF}
	{E}_{\text{H}}[{n_i}, n - n_i] \equiv \frac{1}{2} \int \int \frac{{n}_i({\bf x})[ {n}({\bf x}{'}) - {n}_i({\bf x}{'} )]}{|{\bf x} - {\bf x}{'}|} \mathrm{d} {\bf x} \mathrm{d} {\bf x}{'} .
\end{equation}
The single-orbital energy functional of Eq. (\ref{eqn:HODD-SIF}) is by construction free of the one-electron self-interaction error, Eq. $(\ref{eqn:EHsi})$. The orbital-density-dependent functional of Eq. (\ref{eqn:HODD-SIF}) is by inspection the integral (summation) of terms linear with respect to the orbital density ${n}_i({\bf x})$. Alternatively, the linearity of Eq. (\ref{eqn:HODD-SIF}) is evident from the functional derivative:
\begin{equation}
	\label{eqn:HODD-FSIF}
	\frac{\delta {E}_{\text{H}}[{n_i}, n - n_i]}{\delta {n_i}({\bf x})}  = \int \frac{ {n}({\bf x}{'}) - {n}_i({\bf x}{'} )}{|{\bf x} - {\bf x}{'}|} \mathrm{d} {\bf x}{'} ,
\end{equation}
%\label{eqn:HODD-FSIE}
%\frac{\delta {E}_\text{Hsi }[{n_i}]}{\delta {n_i}({\bf x})} &= \int \frac{{n}_i({\bf x}{'} )}{|{\bf x} - {\bf x}{'}|} \dif {\bf x}{'}
as from the definition of total density $n$ given by Eq. (\ref{eqn:rhodef-total}) the right side is constant with respect to variation of $n_i$. Thus, the Hartree potential associated to the energy functional of Eq. (\ref{eqn:HODD-SIF}) is:
\begin{equation}
	\label{eqn:OFWdef}
	 {{v}_{\text{H}}[{n - n_i}]}({\bf x}) \equiv \frac{\delta {E}_{\text{H}}[{n_i}, n - n_i]}{\delta {n}_i({\bf x})} ,
\end{equation}
which from Eq. (\ref{eqn:HODD-FSIF}) is a functional independent of the orbital density $n_i$. Similarly, with the quantum-linearity of Eq. (\ref{eqn:QL}) the exchange-correlation is constrained to the linear form:
\begin{equation}
	\label{eqn:SIFxc}
	{{v}_{\text{xc}}[{n - n_i}]}({\bf x}) \equiv \frac{\delta {E}_{\text{xc}}[{n_i}, n - n_i]}{\delta {n}_i({\bf x})} ,
\end{equation}
to exclude electron self-interaction and self-exchange energies. The KS total energy functional of Eq. (\ref{eqn:KSEnergy}) is generalized to the set of energy functionals $\{E[{n_i}, n - n_i]\}$ for single-electron orbitals with density $n_i$:
\begin{align}
	\label{eqn:KSEnergy-SIF}
	E[{n_i}, n - n_i] &= T_{\text{s}}[n_i] + E_{\text{H}}[{n_i}, n - n_i] 
    + E_{\text{xc}}[{n_i}, n - n_i] \nonumber \\
    &\quad+ \int v({\bf x})n_i({\bf x}) \mathrm{d} ({\bf x}) ,
\end{align}
The kinetic and external potential terms are functionals of $n_i$ and are without self-interaction.

Following the general approach of \citeauthor{kohnSelfConsistentEquationsIncluding1965}, we postulate (approximate) non-interacting orbitals $\{ \varphi_i \}$ that give the Levy ground-state energy, where each orbital satisfies a one-particle Schr\"{o}dinger equation. These equations are obtained by constrained minimization of the time-independent energy functional:
%\begin{widetext}
\begin{align}
	\label{eqn:LFN-SIF}
	 &\frac{\delta}{\delta \varphi^*_i} \left \{ {E}[{n_i}, n - n_i] - \sum_{ij}^N \varepsilon_{ij}[n - n_i] \right. \nonumber \\
     &\qquad\qquad\qquad \times \left. \left [ \int \varphi_i^*({\bf x}) \varphi_j({\bf x}) \mathrm{d} {\bf x} - \delta_{ij} \right ] \right \} = 0 ,
\end{align}
%\end{widetext}
where $\{ \varphi_i \}$ are orthonormal vectors and the Lagrange multipliers $\varepsilon_{ij}[n - n_i] = \varepsilon^*_{ij}[n - n_i]$ are constant with respect to variation of $\varphi^*_i$. Proceeding as for the KS-equations gives the equations:
\begin{align}
	\label{eqn:eks-SIF}
	&\left \{ \hat{t} + v({\bf x}) + v_{\text{H}}[n - n_i]({\bf x}) + v_{\text{xc}}[n - n_i] ({\bf x}) \right \}
    \varphi_i [{n_i}]({\bf x}) \nonumber \\
	&\qquad = \varepsilon_i [{n}-n_i] \varphi_i [{n_i}]({\bf x}) ,
\end{align}
using the energy functional of Eq. (\ref{eqn:KSEnergy-SIF}), and where Hermitian $\varepsilon_{ij}$ have been diagonalized with a unitary transformation. The orbitals $\varphi_i [{n_i}]$ are functionals of the orbital densities $\{n_i\}$ given by the constrained-search for a ground-state energy of Eqs. (\ref{eqn:LevyF}--\ref{eqn:LE0}). Note the constrained form of Eqs. (\ref{eqn:eks-SIF}) are linear in $\varphi_i$ and thus free of electron self-interaction. The KS effective potential of Eq. (\ref{eqn:KSeff}) becomes:
\begin{equation}
	\label{eqn:eff}
	v_{\text{eff}} [{n}-n_i] \equiv v + v_{\text{H}} [{n}-n_i] + v_{\text{xc}} [{n}-n_i] .
\end{equation}
Note that for one electron ${n}({\bf x}) = {n}_{i}({\bf x})$, and the Hamiltonian of Eq. (\ref{eqn:eks-SIF}) reduces to that of a particle in the external potential $v({\bf x})$. Thus, the latter equations are one-particle self-interaction free. We note that the potential $v_{\text{xc}} [{n}-n_i]$ of Eq. (\ref{eqn:eff}) differs from the approximate form included in the PZ-SIC effective potential of Eq. (\ref{eqn:pzsicpot}). %The approximate exchange energy typically exceeds the correlation energy, and has a non-linear functional dependence on the density.

More generally, assuming fractional occupation numbers $f_i \in [0,1]$ in Eq. (\ref{eqn:rhodef}), the partial derivative of the energy eigenvalue for the linear Eqs. (\ref{eqn:eks-SIF}) is:
\begin{equation}
	\label{eqn:koopmans-odd}
	\left . \frac{\partial \varepsilon_{i}[{{n} - {n}_i}]}{\partial f_{i}} \right |_{\bf f} = 0 ,
\end{equation}
using the definition of ${n}$ given by Eq. (\ref{eqn:rhodef-total}). This result gives the generalized Koopmans condition (GKC) of Eq. (\ref{eqn:koopmans}). That is, the energy of an orbital $\varphi_i$ does not vary with its own fractional occupation $f_i$ \cite{nguyenKoopmansCompliantSpectralFunctionals2018}. Hence, the total system energy is a weighted sum of the energy eigenvalues:
\begin{equation}
	\label{eqn:Etoteigen}
	{E}[\{ {n}_i \}] \equiv  \sum_{i} f_i \varepsilon_{i}[{{n} - {n}_i}] .
\end{equation}
Assuming that for any set of occupation numbers $\{ f_i \}$ there exists a minimum-energy solution of Eq. (\ref{eqn:eks-SIF}) with orbitals $\{ \varphi_i \}$, define the function:
\begin{equation}
\label{eqn:Elineardefn}
{E}(f_1,f_2, \ldots) \equiv {E}[\{ {n}_i \}] ,
\end{equation}
using Eq. (\ref{eqn:rhodef}). The first partial derivative of Eq. (\ref{eqn:Elineardefn}) is:
\begin{equation}
\label{eqn:Janak-odd}
\left . \frac{\partial {E}(f_1,f_2,\ldots)}{\partial f_i} \right |_{\bf f} = \varepsilon_i[{{n} - {n}_i}] ,
\end{equation}
which is a constrained form of Janak's theorem, Eq. (\ref{eqn:Janak}). The second derivative of Eq. (\ref{eqn:Elineardefn}) recovers the GPWL condition of Eq. (\ref{eqn:GPWL}) which has zero curvature for each orbital. The GKC and GPWL conditions Eqs. (\ref{eqn:GPWL}, \ref{eqn:koopmans}) are satisfied by Koopmans-compliant functionals that demonstrate accurate properties for a diverse range of materials \cite{schubertPredictingElectronicScreening2024}. As for the PZ-SIC and Koopmans methods, it may be expected that localized basis functions be required to efficiently achieve self-consistent solutions of the orbital-density-dependent Eqs. (\ref{eqn:eks-SIF}) (see also sections \ref{ss:ODD} and \ref{ss:compresults}).

The non-physical electron self-interaction energy given by Eq. (\ref{eqn:EHsi}) shows a quadratic dependence on the fractional occupation $f_{i}$, with the latter defined by Eq. (\ref{eqn:rhodef}). In addition, the Hartree self-interaction energy of Eq. (\ref{eqn:KSEFH}) increases with orbital localization, i.e., for ${\bf r} \rightarrow 0$. In agreement with these results the KS Hartree energy has been shown to contribute delocalization error with an  approximately quadratic dependence on $f_{i}$ \cite{cococcioniStudySelectedIron2002, cococcioniLinearResponseApproach2005, haitDelocalizationErrorsDensity2018, ColonnaN_2018, colonnaKoopmansSpectralFunctionals2022, linscott2023koopmans}.

The QL-DFT is closely related to but distinct from the Koopmans functionals. Both QL-DFT and the Koopmans integral (KI) functional are compliant with the GKC and GPWL condition. The KI functional applies the GKC to a KS functional, by the variational energy minimization in a frozen orbital approximation, where the latter is compensated by  screening effects to relax the orbitals \cite{linscott2023koopmans, schubertPredictingElectronicScreening2024}. The KIPZ functional applies the PZ-SIC to the KI functional, whereby  orbital-density-dependent corrections are applied to the approximate KS-functionals, which leaves the KS-equations formally intact. In contrast, the QL-DFT approach solves Eqs. (\ref{eqn:eks-SIF}), which are themselves compliant with the GKC and GPWL conditions. In the following sections we compare QL-DFT fundamental band gaps with those of the KIPZ functional, and other methods from the literature.

\begin{table*}[!htb]
\caption{\label{table:bgkipz}Comparison of QL-DFT fundamental band gaps (eV) with the WFL-SIC and KIPZ (GKC) functional corrections to KS-DFT at zero-temperature \cite{shindeImprovedBandGaps2020, degennaroBlochsTheoremOrbitaldensitydependent2022}. The band gaps compare with the experimental values corrected for zero-point renormalization (Exp+ZPR). The mean error (ME) and mean absolute error (MAE) with respect to Exp+ZPR values are also shown. 
} % end caption
% Band gaps: gadi:/g/data/jh25/ji2211/
% C-dia: qe/qldft/v7d0/addk-15726/c-dia/ol-oncv, op-oncv, ops-oncv, README_bandgap
% Si-dia, ...: qe/qldft/v7d0/scf-bands/si-dia/ol, op, ops, README_bandgap
\begin{ruledtabular}
\begin{tabular}{lccccccccc}
	&\multicolumn{3}{c}{KS-DFT } &
	\multicolumn{3}{c}{QL-DFT} &
    \multicolumn{1}{c}{WFL-SIC\footnotemark[1]} &
	\multicolumn{1}{c}{KIPZ} &
	\multirow{2}{*}{Exp+ZPR} \\ 
	&LDA&PBE&PBEsol&LDA&PBE&PBEsol&PBE&PBE& \\ 
    \hline
	C-dia & $4.15$ & $4.20$ & $4.06$ & $5.42$ & $5.59$ & $5.43$ & $5.3$ & $6.58$\footnotemark[2] & $5.80$\footnotemark[4]  \\
    Si-dia & $0.48$ & $0.57$ & $0.46$ & $1.17$ & $1.36$ & $1.23$ & $1.7$ & $1.19$\footnotemark[2] & $1.23$\footnotemark[4] \\
    BN-zb & $4.41$ & $4.52$ & $4.33$ & $7.18$ & $7.66$ & $7.38$ & $-$ & $6.73$\footnotemark[2] & $6.65
$\footnotemark[4]  \\
    BP-zb & $1.18$ & $1.23$ & $1.12$ & $1.94$ & $2.11$ & $1.95$ & $2.5$ & $2.27$\footnotemark[3] & $2.36$\footnotemark[5]  \\
    SiC-zb & $1.33$ & $1.36$ & $1.24$ & $3.42$ & $3.74$ & $3.49$ & $4.7$ & $2.40$\footnotemark[3] & $2.58
$\footnotemark[4]  \\
    AlP-zb & $1.46$ & $1.59$ & $1.45$ & $3.05$ & $3.40$ & $3.25$ & $4.0$ & $2.60$\footnotemark[3] & $2.55$\footnotemark[4]  \\
    ME & $-1.36$ & $-1.28$ & $-1.42$ & $0.17$ & $0.45$ & $0.28$ & $0.74
$ &$0.10$ & $ $ \\    
    MAE & $1.36$ & $1.28$ & $1.42$ & $0.45$ & $0.60$ & $0.47$ & $0.94$ &$0.20$ & $ $ \\ 
\end{tabular}	
\end{ruledtabular}
\footnotetext[1]{Ref.~\onlinecite{shindeImprovedBandGaps2020}}
\footnotetext[2]{Ref.~\onlinecite{degennaroBlochsTheoremOrbitaldensitydependent2022}}
\footnotetext[3]{Ref.~\onlinecite{nguyenKoopmansCompliantSpectralFunctionals2018}}
\footnotetext[4]{Ref.~\onlinecite{MiglioA_2020, engelZeropointRenormalizationBand2022}}
\footnotetext[5]{Ref.~\onlinecite{wooBPSynthesisProperties2016, mageratZeropointBandGap}}
\end{table*}
\begin{table*}[!htb]
\centering
\caption{\label{table:gkc-comp}Comparison of QL-DFT fundamental band gaps (eV) with other computational methods from the literature. The ME and MAE refer to experimental values (Exp+ZPR).
} % end caption
\begin{ruledtabular}
\begin{tabular}{lccccccc}
    &\multicolumn{4}{c}{KS-DFT\footnotemark[1]} &
	\multicolumn{1}{c}{QL-DFT} &
	\multicolumn{2}{c}{MBPT } \\ 
	&mBJLDA&HLE16&HSE06&sLOSC&LDA&$G_0W_0$\footnotemark[2]&QS$G\tilde{W}$\footnotemark[3]\\ 
    \hline
	C-dia  &$4.92$ & $4.55$ & $5.26$ & $4.71$& $5.42$ & $5.49$--$5.96$ & $5.64$--$5.90$ \\
    Si-dia & $1.15$ & $1.38$ & $1.17$ &$1.57$ & $1.17$ & $1.03$--$1.26$ & $1.08$--$1.30$ \\
    BN-zb & $5.80$ & $5.38$ & $5.76$ &$6.01$ & $7.18$ & $6.04$--$6.71$ & $6.30$--$6.67$ \\
    BP-zb & $1.85$ & $1.85$ & $1.98$ &$2.05$ & $1.94$ & $1.95$--$2.18$ & $2.05$ \\
    SiC-zb & $2.25$ & $2.06$ & $2.23$ &$1.96$ & $3.42$ & $2.23$--$2.52$ & $2.35$--$2.53$  \\
    AlP-zb & $2.31$ & $2.75$ & $2.30$ &$2.97$ & $3.05$ & $2.25$-$2.49$ & $2.42$--$3.10$  \\
    ME & $-0.48$ & $-0.53$ & $-0.41$ &$-0.32$ & $0.17$ & $-0.19$\footnotemark[4] & $-0.08$\footnotemark[4] \\     
    MAE & $0.48$ & $0.65$ & $0.41$ &$0.57$ & $0.45$ & $0.19$\footnotemark[4] & $0.15$\footnotemark[4] \\ 
\end{tabular}	
\end{ruledtabular}
% KS functionals:
\footnotetext[1]{Ref.~\onlinecite{tran2017, TranF_2009, vermaHLE16LocalKohn2017, heydHybridFunctionalsBased2003, mahlerLocalizedOrbitalScaling2022}} 
% G_0W_0:
\footnotetext[2]{Ref.~\onlinecite{nguyenKoopmansCompliantSpectralFunctionals2018, azizi2025, renAllelectronPeriodic$G_0W_0$2021, salas-illanesElectronicStructureRepresentative2022, chenAccurateBandGaps2015, kutepovSelfconsistentSolutionHedins2017, surhQuasiparticleEnergiesCubic1991, grossmannManybodyPerturbationTheory2026}} 
% QSGW~:
\footnotetext[3]{Ref.~\onlinecite{chenAccurateBandGaps2015, kutepovSelfconsistentSolutionHedins2017, cunningham$mathrmQSGstackrelifmmodeHatelseFiW$2023, grossmannManybodyPerturbationTheory2026}} 
\footnotetext[4]{Based on the mid-value of each band gap range}
\end{table*}
% C:\Users\johne\Documents\Documents\PhD\mypapers\ql-dft\figures
\begin{figure*}[!htb]
        \centering
        \includegraphics[height=9.0cm]{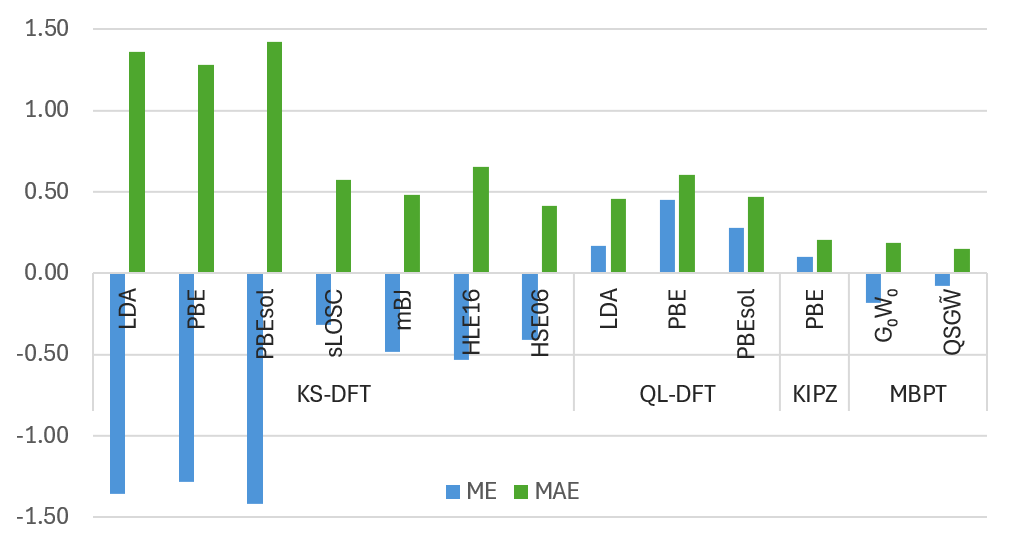} \\    
	\caption{\label{fig:BAR} The mean error (ME, blue) and mean absolute error (MAE, green) of the fundamental band gap with respect to experimental values (Exp+ZPR). The KS-DFT and advanced KS-DFT functionals (left) are compared with: QL-DFT (center); the KIPZ functional and MBPT (right). The QL-DFT band gaps are given for non-polarized spin at zero temperature. %The experimental gaps have been corrected for zero-point renormalization (ZPR). 
    }
\end{figure*}
% /g/data/jh25/ji2211/qe/qldft/v7d0/addk-15726/c-dia
% /g/data/jh25/ji2211/qe/qldft/v7d0/scf-bands/si-dia, ...
% Gnuplot: C:/cygwin64/home/johne/mypapers-cyg/dft-sie-cyg/bands/csibn, bpsicalp
\begin{figure*}[!htb]
	\centering
    \input{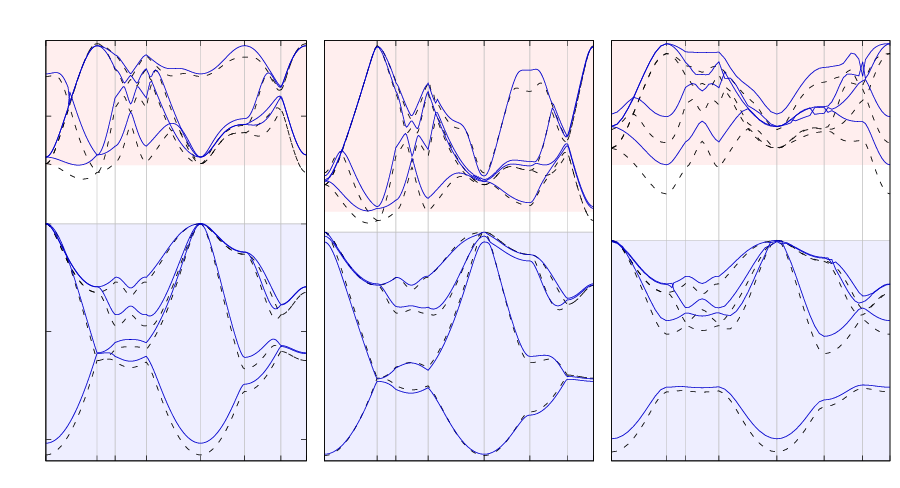}
    \vspace{12pt}
    \input{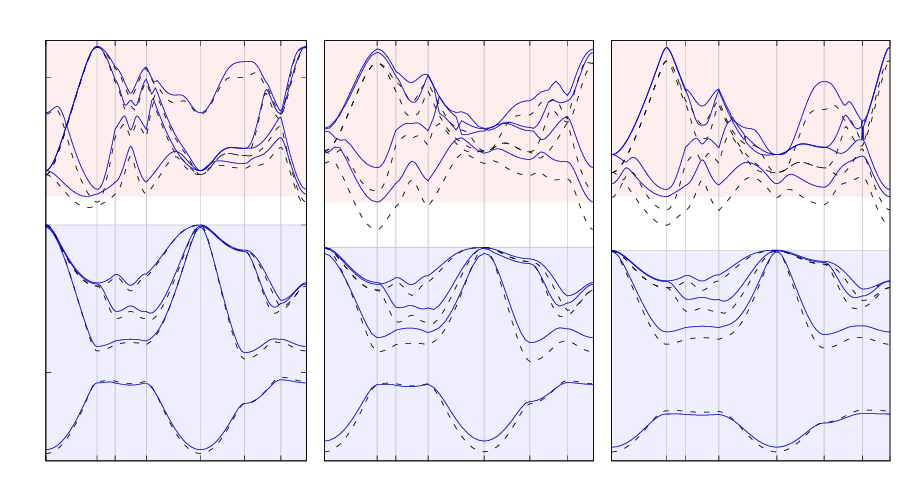}
	\caption{\label{fig:SIBNBP}Comparison of the KS-DFT (black, dashed) and QL-DFT (blue) zero-temperature electronic band structures of: (a) \ce{C}; (b) \ce{Si}; (c) \ce{BN}; (d) \ce{BP}; (e) \ce{SiC}; and (f) \ce{AlP}, with non-polarized spin in the local density approximation (LDA). The zero energy has been set to the valence band maximum (VBM) energies.
    }
\end{figure*}

% What was done
\section{\label{ss:compdetails}Computational details}
The quantum-linear (QL) self-consistent field (SCF) Eqs. (\ref{eqn:eks-SIF}) have been implemented in the \texttt{Quantum ESPRESSO} software package \cite{giannozziAdvancedCapabilitiesMaterials2017, giannozziQUANTUMESPRESSOModular2009}. Scalar-relativistic PseudoDojo Optimized Norm-Conserving Vanderbilt (ONCV) pseudopotentials were used \cite{  hamannOptimizedNormconservingVanderbilt2013, vansettenPseudoDojoTrainingGrading2018}. Functionals comprise the local density approximation (LDA) \cite{perdewAccurateSimpleAnalytic1992}, Perdew-Burke-Ernzerhof (PBE) \cite{Perdew_1996} exchange-correlation, and PBE with parameters for bulk solids (PBEsol) \cite{perdewRestoringDensityGradientExpansion2008a}. The modification of \texttt{Quantum ESPRESSO}  has been implemented for serial execution using conjugate gradient (CG) diagonalization of the Hamiltonian for the valence electrons, and unpolarized spin of the orbitals without smearing \cite{giannozziAdvancedCapabilitiesMaterials2017}. The pseudopotential of the core electrons \cite{ hamannOptimizedNormconservingVanderbilt2013,  vansettenPseudoDojoTrainingGrading2018} remain unchanged. 

The crystal structures and lattice parameters were obtained from the Inorganic Crystal Structure Database (ICSD) \cite{RuhlS_2019}. The unit cells were relaxed using (unmodified KS-DFT) Quantum ESPRESSO with the PBEsol functional to an electron convergence threshold of $10^{-14}$ Ry \cite{csonkaAssessingPerformanceRecent2009, perdewRestoringDensityGradientExpansion2008a, zhangPerformanceVariousDensityfunctional2018a, haoLatticeConstantsSemilocal2012, shaoAssessingPerformanceExchangecorrelation2023, ingallAccurateEfficientComputation2024}, and the results analyzed using XCrySDen software \cite{kokaljXCrySDenNewProgram1999}. A kinetic-energy energy cutoff of $100$ Ry (dual $4$ for the charge densities) was used for all results shown in this paper, with an SCF convergence threshold of $10^{-7}$ Ry. 

The k-point convergence differs qualitatively and quantitatively from that of KS-DFT. As compared with KS-DFT, localized orbitals are expected to result from the subtraction of delocalization error. To facilitate k-point convergence, the Oxidation-State Constrained (OS-C) DFT plug-in to Quantum ESPRESSO provided L\"{o}wdin orthogonalized atomic orbitals (without imposing occupation constraints) \cite{lowdinNonOrthogonalityProblemConnected1950, kuOxidationStateConstrainedDensity2019}. The resulting k-point convergence is shown in Table \ref{table:kpoint} of the Appendix. For the present implementation, (delocalized) KS-orbitals provide a suitable starting-point for a QL-DFT calculation. The k-points in reciprocal space formed a Monkhorst-Pack (MP) mesh  \cite{monkhorstSpecialPointsBrillouinzone1976}, with a path of high-symmetry k-points provided by the \texttt{Seekpath} library \cite{hinumaBandStructureDiagram2017a, togoSpglibSoftwareLibrary2024, ase-paper}. The energy eigenvalues along band-structure paths were determined by including these k-points in the SCF calculation with zero weight. The latter method was previously developed for hybrid functionals, to remove the need for a subsequent non-SCF calculation that assumes a fixed potential.

\section{Results and Discussion}
\label{ss:compresults}
The fundamental band gap has been computed for the materials listed in Table \ref{table:kpoint} of the Appendix. 
The present implementation of quantum-linear (QL) DFT requires a higher density of the k-point grid for convergence as is typically required for KS-DFT \cite{kratzerBasicsElectronicStructure2019a}. Bulk carbon with the $Fd\bar{3}m$ (227) space group required the largest number of k-points for precise convergence (see Table \ref{table:kpoint}). These results are in accord with an expected spatial localization of orbitals on subtraction of the (repulsive) self-interaction given by Eq. (\ref{eqn:EHsi}). 
The computed fundamental band gaps of the bulk solids are shown in Table \ref{table:bgkipz}. 
% Table 1 above
The band gaps are compared with those computed with the WFL-SIC and KIPZ functional corrections to KS-DFT \cite{degennaroBlochsTheoremOrbitaldensitydependent2022}, and experimental values corrected for zero-point renormalization. The mean error (ME) and mean absolute error (MAE) assign equal weight to the error for each material. We note the mean absolute percentage error (MAPE) is increased (decreased) for narrow- (wide-) gap insulators \cite{grossmannManybodyPerturbationTheory2026}. A mean valued of the QL-DFT band gap for each solid was calculated using the minimum number of grid k-points (and greater) shown in Table \ref{table:kpoint}. For QL-DFT, the LDA functional shows the greatest accuracy (ME) and precision (MAE) as compared to QL@PBE and QL@PBEsol. The PBE and PBEsol gradient corrections to QL@LDA systematically increase the gap for each solid, with the PBE correction larger than PBEsol. That the gradient corrections increase the ME and MAE can  be attributed to slight average overestimates ($\sim 0.3$ eV) of the band gaps by QL@LDA. The QL@PBE method shows increased ME and MAE as compared to the KIPZ@PBE functional, the former computed without spin polarization and smearing. 

The QL-DFT fundamental band gaps are compared with values from the literature in Table \ref{table:gkc-comp}.
% Table 2 above
These band gaps have been reported previously at the meta-GGA and hybrid level, \emph{viz.} using the mBJLDA, HLE16, HSE06 and sLOSC exchange correlation functionals \cite{borlidoExchangecorrelationFunctionalsBand2020,tran2017, TranF_2009, vermaHLE16LocalKohn2017, heydHybridFunctionalsBased2003, mahlerLocalizedOrbitalScaling2022}. In addition, the WOT-SRSH functional gaps for \ce{C}, \ce{Si} and \ce{AlP} are $5.7$, $1.1$ and $2.6$ eV, respectively \cite{wingBandGapsCrystalline2021}. For MBPT, the gaps have been reported previously using the $G_0W_0$ approximation and the quasiparticle self-consistent $GW$ method with vertex corrections (QS$G\tilde{W}$) \cite{nguyenKoopmansCompliantSpectralFunctionals2018, azizi2025, renAllelectronPeriodic$G_0W_0$2021, salas-illanesElectronicStructureRepresentative2022, chenAccurateBandGaps2015, kutepovSelfconsistentSolutionHedins2017, surhQuasiparticleEnergiesCubic1991, kutepovSelfconsistentSolutionHedins2017, cunningham$mathrmQSGstackrelifmmodeHatelseFiW$2023}. Note that the $G_0W_0$ gaps shown in Table \ref{table:gkc-comp} generally increase in accuracy with the use of all-electron, full-potential codes and Slater-type or high-energy localized orbitals \cite{renAllelectronPeriodic$G_0W_0$2021, salas-illanesElectronicStructureRepresentative2022}. The $G_0W_0$ and QS$GW$ (without vertex corrections) approximations can underestimate and overestimate, respectively, the fundamental band gap of semiconductors and insulators \cite{golzeGWCompendiumPractical2019}.
The QL@LDA fundamental band gaps of Tables \ref{table:bgkipz} and \ref{table:gkc-comp} show close agreement with experimental values and the approximations of MBPT. Remarkably, the QL@LDA has closed the band gaps with a $87.5$\% ($66.9$\%) decrease in the  ME (MAE) as compared with KS@LDA. For the $G_0W_0$ approximation of MBPT, QL@LDA shows a $10.5$\% decrease in the ME with a $137$\% increase in the MAE. A band gap for silicon of $1.229(8)$ eV for QL@PBEsol shows exceptional agreement with the ZPR-corrected experimental value of $1.228$ eV.

% Figure 1 above
Figure \ref{fig:BAR} compares the (experimental) ME and MAE of the band gaps from Tables \ref{table:bgkipz} and \ref{table:gkc-comp}. There is a large reduction of the KS-DFT band gap errors, particularly for QL@LDA. The QL@LDA errors compare well with the advanced KS-DFT functionals, including the hybrid functionals with their associated computational cost. The ME of QL@LDA shows an accuracy comparable to the `single shot' $G_0W_0$ method of MBPT.

% Figure 2 above
Figure \ref{fig:SIBNBP} shows the QL@LDA electronic band structures for the solids. The QL@LDA indirect band gaps open between the $\Gamma$ and X high symmetry points, as compared to KS@LDA. The band structures for \ce{Si}, \ce{C}, and \ce{BN} compare qualitatively with the KIPZ band structures reported by \citeauthor{degennaroBlochsTheoremOrbitaldensitydependent2022}\cite{degennaroBlochsTheoremOrbitaldensitydependent2022}. \ce{BN} exhibits significant SIE which is especially difficult to describe using KS density-functional approximations \cite{gouldPoisoningDensityFunctional2022}. Figure \ref{fig:SIBNBP} and Table \ref{table:bgkipz} show that QL@LDA opens the \ce{BN} gap by a substantial $2.77$ eV.

\section{Summary and outlook}
In this work the Kohn-Sham equations are constrained to subtract non-physical electron self-interaction exactly, in accordance with generalized Hohenberg-Kohn theorems. The resulting equations are linear for each non-interacting electron, while the system of coupled equations remain implicitly non-linear. In the absence of self-interaction, the eigenenergy of each orbital is invariant with respect to its occupation (GKC). The total energy is a occupation-number weighted sum of the eigenenergies, which is a piecewise-linear functional of the orbital occupation (GPWL). The Hartree component of the self-interaction error, Eq. (\ref{eqn:EHsi}), shows a quadratic dependence on fractional occupation, which is increased for localized atomic orbitals. 
The converged solutions require an increased density of the k-point grid (Table \ref{table:kpoint}) as compared with those of KS-DFT, in accord with the subtraction of (repulsive) orbital self-interaction with consequent orbital localization. This approach when applied in conjunction with LDA and GGA (PBE, PBEsol) exchange correlation functionals yield fundamental band gaps with an accuracy comparable to that of MBPT (Table \ref{table:gkc-comp}). Preliminary numerical results support the hypothesis that the KS effective potential can be constrained to a functional dependence on orbital densities to remove orbital self-interaction error (SIE) from the Kohn-Sham equations. 

In conclusion, implementation of the exact GPWL condition in DFT codes involves replacement of the approximate Kohn-Sham potential $v_{\text{s}}[{n}]$, by an orbital-density-dependent potential $v_{\text{eff}}[{n - n_i}]$. Self-consistent minimization of the total system energy is realized by constrained variation of the orbitals and their fractional occupations. These results can facilitate the highly accurate simulation of material properties in physics and quantum chemistry, in particular the rational design of sustainable materials and processes.

\begin{acknowledgments}
JEI would like to thank Associate Professor Vicki J. Keast for her guidance and review of early versions of the manuscript; Associate Professor Karen L. Livesey for her comments and suggestions; and the University of Newcastle Computational Chemistry group for their suggestions. We are grateful to Dr Edward Linscott and Dr Nicola Colonna whose suggestions included close comparison with the PZ-SIC. JEI would also like to thank participants at the following conferences for their valued comments and suggestions: Psi-k 2025, Lausanne, Switzerland; 2025 Gordon Godfrey Workshop, The University of New South Wales, Australia; 2025 Australian Institute of Physics (AIP) Summer Meeting, The University of Wollongong, Australia; and the 2026 AIP 48th Annual Condensed Matter and Materials Meeting, in conjunction with the 6th Asia-Pacific Conference on Condensed Matter Physics, Charles Sturt University, Wagga Wagga, Australia. JEI gratefully acknowledges financial support from an Australian Government Research Training Program Scholarship (RTP). We gratefully acknowledge the National Computational Infrastructure (NCI) supported by the Australian Government for computing resources and support services.
\end{acknowledgments}

\section{Data availability}
The data that support the findings of this study were generated by numerical simulations. The source code and parameters used to generate the simulations are openly available \cite{ingallqldftdata2026}.

%\newpage
% Specify following sections are appendices. Use \appendix* if there
% only one appendix.
\appendix*
\section{Crystal structures and k-point convergence}
\label{sec:appendixa}
%The crystal structures and k-point convergence are shown in Table \ref{table:kpoint}.
% Archive for all tables and figures on res-dm.newcastle.edu.au:~c3337930/ARCHIVE/gadi/
% Carbon: qe/koop/carbon/README_bandgap
%% LDA: gdata/jh25/ji2211/qe/koop/carbon/k20 k30 (+scratch/) k40 k50v2
%% PBE: pbe/k40
%% PBEsol: pbesol/k40
% Silicon: qe/koop/si/README_bandgap
%% LDA: gdata/jh25/ji2211/qe/koop/si/icsd-path/k60v2 k70 (+scratch/) k80 k90
%% PBE: pbe/k80 (+scratch/)
%% PBEsol: pbesol/k80
% BN: gdata/jh25/ji2211/qe/koop/bn/README_bandgap
%% LDA: gdata/jh25/ji2211/qe/koop/bn/ec100/k50v1 k60v1 k70v1 k80v2 k90v1_nc
%% PBE: pbe/k70v1
%% PBEsol: pbesol/k70v1
% BP: gdata/jh25/ji2211/qe/koop/bp/zb/a4d534/README_bandgap
%% LDA: k50v1_crash  k60v1  k70v1  k80v1 k90v1
%% PBE: pbe/k80v1
%% PBEsol: pbesol/k80v1
% AlP: /g/data/jh25/ji2211/qe/koop/extended/xcv/alp/README_bandgap
%% LDA: alp/addk/k24v2
%% PBE: alp/addk/pbe/k24v2
%% PBEsol: alp/addk/pbesol/k24v2
\begin{table}[!htb]
\caption{\label{table:kpoint}Crystal structures \cite{RuhlS_2019, bindzusExperimentalDeterminationCore2014, rossEmpiricalAssessmentTwodimensional2014, willChargeDensityChemical1986, xiaBPMegabarPressures1993a, kawamuraSILICONCARBIDECRYSTALS1965, richmanVaporPhaseGrowth1968}, computed lattice parameter ($a$), and the number of k-points in the reciprocal-space grid for band gap convergence using the QL@PBEsol method. The standard error of the mean (SEM) band gap values ($E_g$) for increasing grid size are shown using the QL@LDA, QL@PBE and QL@PBEsol density functionals, respectively. %The present implementation requires a larger number of k-points for SCF convergence as compared with KS-DFT.
} % end caption
\begin{ruledtabular}
\begin{tabular}{lcccccc}
    &&&&\multicolumn{3}{c}{ SEM $E_g$ (meV)} \\
    &Structure&$a$ ($\accentset{\circ}{\text{A}}$)& k-point grid & LDA & PBE & PBEsol \\ 
    \hline
    C & diamond & $3.55835$ & $48^3$\footnotemark[1] & $14$ & $8$ & $6$ \\ % ICSD 190653; Shishkin & Kress 2007
    Si& diamond & $5.43057$ & $20^3$ & $16$ & $6$ & $8$ \\ % ICSD 257794
    BN& zincblende & $3.60744$ & $16^3$ & $2$ & $4$ & $5$ \\ % ICSD; Shishkin & Kress 2007
    BP& zincblende & $4.52060$ & $20^3$ & $1$ & $1$ & $1$ \\ %  ICSD 602964
    SiC& zincblende & $4.35783$ & $20^3$ & $4$ & $8$ & $5$ \\ % 
    AlP& zincblende & $5.46872$ & $20^3$ & $0$ & $1$ & $1$\\ %  
\end{tabular}	
\end{ruledtabular} 
\footnotetext[1]{Band-path interpolation was not given (nor required) for these high k-point grid densities}
\end{table}

%\newpage
%\bibliography{apssamp}% Produces the bibliography via BibTeX.
%\bibliography{kse,thesis}
\bibliography{kse}
\end{document}